\documentstyle[prl,aps]{revtex}
\begin{document}
\draft
\tighten
\twocolumn[\hsize\textwidth\columnwidth\hsize\csname
@twocolumnfalse\endcsname \title{Universalities in vortex  transport
at the melting transition} \author{A.G.Rojo$^a$, A.K.Sood$^b$ and
C.A.Balseiro$^c$}\address{$^a$Department of Physics, The University of
Michigan, Ann Arbor, MI 48109-1120} \address{$^b$Department of
Physics,Indian Institute of Science, Bangalore 560 1012, India }
\address{$^c$Centro At\'omico Bariloche and Instituto Balseiro,
8400-Bariloche, Argentina} \maketitle

\begin{abstract}
We consider the jump in resistance at the melting transition, which is
experimentally observed to be constant, independent of magnetic field
(vortex density). We present an explanation of this effect based on
vortex cuttings, and universalities of the 
  structure factor at the freezing transition (the
Hansen--Verlet criterion).
\end{abstract}

\pacs{PACS numbers: 47.32.Cc,64.70Dv,74.60.Ge}
\vskip2pc] \narrowtext

The subject of vortex dynamics in high Tc superconductors is of
topical interest \cite{crabtree,blatter}. In recent years, new aspects of the
phase diagram of the vortex lattice have been elucidated through
experiment. Perhaps one of the salient recent experimental findings is
the first order melting transition of the vortex lattice, first
observed in transport measurements \cite{safar} and later confirmed in
a number of equilibrium properties\cite {bariloche}. It is now well
accepted that in clean BiSrCuO and in untwined YBaCuO samples the
melting transition of the vortex lattice for an applied magnetic field
parallel to the $c$-axis is first order and that the superconducting
coherence is simultaneously lost in all directions at the melting
temperature. A striking regularity is observed in the transport
experiments: for different values of the applied magnetic field, which
is proportional to the areal density of vortices, the in-plane
resistivity $ \rho_{ab}(T)$ decreases with decreasing temperature
until it jumps (essentially) discontinuously to zero at the melting
temperature $T_m$, with $\rho_{ab}(T_m)$ {\em independent } of field.
Moreover, recent studies of the melting transition in YBaCuO show that
the jump in the out-of-plane resistivity $\rho_c$ is also field
independent.

This universality in the jump in resistance at the melting temperature
constitutes, we argue, an evidence of general universalities in the
structure factor $S(q)$ of the liquid phase at the melting point as
first pointed out by Hansen and Verlet for normal
liquids\cite{hansen}.

Verlet's criterion for freezing  is the counterpart of
Lindeman's criterion for melting. Verlet observed that the rescaled
curves of $S(q)$ are nearly identical for a variety of liquids that
freeze to the same crystalline structure. The only scale is $q_m$, the
wave vector at the first peak in $S(q)$, with $S(q_m)\simeq 2.85 $.

While Lindeman's criterion tells us a way of anticipating melting by
looking at mean square displacements of the atoms in the solid phase
\cite{ziman}, Verlet's criterion provides a way of predicting freezing
by looking at the structure factor in the liquid phase. Ramakrishnan
and Yussouf\cite{tvr} have offered a theoretical explanation of the
criterion based on a phenomenological order parameter theory.

More recently  L\"owen {\it et al.}\cite{lowen} made the empirical
observation of a {\em dynamical} criterion for freezing of colloidal
liquids. For these systems the ratio of the long-time and short-time
diffusion coefficients of the colloidal particles $D_L/D_S$ has a
universal value close to $0.1$ at the freezing temperature of a
variety of systems. It has been shown that, for the colloidal liquids,
universalities in $S(q)$ at freezing imply universalities in the long
time diffusion constant\cite {indrani,medina}.

In the present paper, we will assume the validity of Verlet's
criterion for vortices. However, we wish to emphasize that there are
still important differences in the the diffusion coefficients of
vortices and colloidal particles which stem from the different
topological properties of these systems. Vortices are extended objects
that can entangle. In the entangled phase diffusion implies cutting
of vortex lines, an effect which not present in suspended particles.
For clean systems we show that vortex cutting, together with Verlet's
criterion, implies a universality in the jump of both $\rho _{ab}$ and
$\rho _c$. We illustrate this by  solving a simple model for
vortex cutting, and show that the resistivity
at the melting temperature is given
 by the normal resistivity and the structure factor only, and thus the
observed universality in the resistivity jump implies the validity of
Verlet's criterion. 

We will first analyze the out of plane resistivity $\rho_c$. The
mechanisms of dissipation when the current is parallel to the applied
field originate in the entanglement of the field--induced vortices and
the thermally generated vortex loops\cite{jagla}. In this entangled
phase, the structure of vortex lines percolates through the sample in
the direction perpendicular to the external field. The result is a
``phantom mesh" of vortex lines that threads the sample in all
directions and where each vortex diffuses by cutting the others. The
topology of the phantom mesh allows different segments of the 
field--induced vortices to diffuse independently. This implies 
that the vortex
motion is uncorrelated in the $c$--direction, as observed in the pseudo DC
transformer experiments\cite{pastoriza}.

When a current flows along the $c$--axis, only those lines oriented in
the direction perpendicular to the $c$--axis will contribute to the
bulk resistivity. The resistivity can be written as \begin{equation}
\rho_c = \left({\frac{\phi_0 }{c}}\right)^2 n_P {\frac{D_P }{k_B T}},
\end{equation} with $n_P$ the density of percolating vortex lines, and
$D_P$ their diffusion coefficient. In the computation of $D_P$ one
needs to include the cutting of the percolating vortex lines with the
field--induced vortices.

We compute the diffusion by considering the random walk of the center
of mass of a vortex line. We introduce a discretization of the random
walk, with step length ${\ell} $, and we call $\tau_i$ the time
elapsed at the $i$--th step. Neglecting vortex cutting all time
steps are equal, with $ \tau_i=\tau_0$, and the bare diffusion
coefficient is given by $ D_0=\ell^2/\tau_0$. This bare diffusion
constant is related to the Bardeen--Stephen\cite{bs,tinkham} viscosity coefficient
$\eta_0= k_B T /D_0 $.

Assume now that there is a fraction $n_c$ of steps, at which vortex
cutting takes place, with a characteristic step time $\tau_{\times}$;
the diffusion coefficient is then
 \begin{equation} D_P={\ell}
^2/\left[n_c \tau_{\times} + (1-n_c) \tau_0 \right]. \end{equation}
The cutting frequency is \begin{equation} {\frac{1}{\tau_{\times} }} =
{\frac{1}{\tau_0 }}e^{ -U_{\times}/k_B T}, \end{equation} with
$U_{\times}$ an effective barrier for cutting\cite{nelson_cut}. In the
above equation we have assumed thermal activation for the vortex
crossings. The Boltzman factor gives the relative probability of
finding a vortex in a crossing configuration.

If  $k_B T <U_{\times}$, and assuming that $n_c$ is of the order
of the density of field--induced vortices $n$, the resistivity is given by

\begin{equation}
\rho_c =\left( {\frac{\phi_0 }{c}}\right)^2 {\frac{1}{\eta_0}} \left
({\frac{ a_0}{\xi _z}}\right)^2 e^ {-U_{\times}/k_B T},  \label{roc}
\end{equation} where $a_0=n^{-1/2}$ is the mean distance between field induced
vortices along the $c$--direction and $\xi _z=n_P^{-1/2}$ is the mean distance
between percolating vortices in the direction perpendicular to the
field. In what follows, we argue that Verlet's criterion implies that
$\rho_c$ is universal (in a weak sense) at the melting temperature.
The argument is two--fold: {\em i}--The universality of the structure
factor $S(q)$ indicates that the scaled correlation lengths are
independent of density at $T_m$, and therefore the factor $\left({a_0
/ \xi _z}\right)^2$ is density independent at freezing. {\em ii}--The
exponential factor in Eq. (\ref{roc}) represents an effective
probability of cutting which will be also universal at melting due to
Verlet' criterion. (See the solution of the 
 simplified one--dimensional model below.)          

We conclude that the out--of--plane resistivity $\rho _c$ is
independent of the vortex density at the melting temperature $T_m$.
This weak universality is a consequence of Verlet's criterion. The
jump in the in--plane resistivity is also independent of the external
field. The resistivity $\rho _{ab}$ can be written as \begin{equation}
\rho _{ab}=\left( {\frac{\phi _0}c}\right) ^2n{\frac D{k_BT}},
\end{equation} with $n=B/\phi _0$  the density of field induced
vortices along the direction of the applied field $B$, $D$ is the
corresponding diffusion coefficient. The
diffusion coefficient $D$  can be calculated  following the out--of--plane
case, with an analogous results. We stress that the diffusion constant is reduced by a factor
proportional to $n$ due to vortex cutting, canceling the prefactor of
$n$ in the above equation.

The arguments presented above are valid for good quality single
crystals for which vortex pinning is weak. Evidence from both
experiments and numerical simulations have shown that intrinsic
disorder destroys the first order transition. For example, upon
irradiation, the magnitude of the jump decreases. In the cleanest YBCO
samples, the jump in $\rho _{ab}$ is about a fifth of the normal
resistivity at the critical temperature. A smaller jump in the
resistivity at the melting transition is an indication of disorder. In
our formulation, a large disorder can be incorporated in a 
modified particle--particle correlation function which will retain
its universal properties as long as the transition remains first order. 
 Consequently, to the extent that Verlet's criterion
remains valid, the universality in the resistity jump is preserved.

Now let us turn to a simplified one--dimensional model of interacting
diffusing particles. The purpose of this discussion is to justify
using a probability of cutting which is universal at melting. We
consider the following Fokker--Planck equation, describing diffusing
interacting particles

\begin{equation}
{\frac{\partial}{\partial t}} P(\{x_i\},t) = \left\{\sum_{i=0}^N
D_i{\frac{ \partial ^2 }{\partial x_i^2}} -\sum_{i=0}^N
{\frac{D_i}{k_BT}} {\frac{ \partial }{\partial x_i}} F_i
\right\}P(\{x_i\},t) , \end{equation}

with $F_i= -\delta_{i,0}\left[f
+ \sum_{j=1}^N V^{\prime}(x_j-x_0)\right] + (1-\delta_{i,0})
V^{\prime}(x_i-x_0)$,
$V$ being an interaction potential between
particle zero and the rest of the $N$ particles. Note that there is no
interaction between particle $i$ and particle $j$ for $i\neq 0\neq j$.
Also, the diffusion constant is $D_0$ for particle $0$, and $D_i \ll
D_0$ for $i\neq 0$. We added a drag force $f$ acting on particle zero,
and computed the resulting mean drift velocity of particle zero
$\langle v\rangle $ by solving the Focker--Planck equation imposing
periodic boundary conditions\cite{risken} on a length $L$. The result
is $\langle v\rangle= (D/k_BT) f$, with

\begin{equation}
{\frac{D}{D_0}}= {\frac{L }{\int_0 ^{L} dx e^{- \sum_{i=1}^NV(x
-x_i)/k_BT} } } {\frac{L }{\int_0 ^{L} d x e^{+ \sum_{i=1}^NV(x
-x_i)/k_BT} }}.  \label{d0} \end{equation}

Note that, since particle zero diffuses faster than the other
particles, we take the configuration of the rest of the particles as
described by the equilibrium distribution in the absence of the force
$f$. Also, if the potential is monotonically decreasing with distance
with $V(0)\gg k_BT$ and the range of the interaction is much smaller
than the inter--particle separation,

\begin{equation}
{\frac 1L}\int_0^Ldxe^{+\sum_{i=1}^NV(x-x_i)/k_BT}\sim {\frac
NL}\lambda e^{V(0)/k_BT},  \label{imp} 
\end{equation}

with $\lambda $
the range of the interaction. We can now identify the integral in Eq.
(\ref{d0}) which involves the negative exponential as a partition
function $Z$, and defining $P_{\times }=[\exp {-V(0)/k_BT}]/Z$, we
obtain \begin{equation} {\frac D{D_0}}=\frac 1{n\lambda }P_{\times }.
\end{equation} In this simplified model $P_{\times }$ represent the
probability of finding particle zero on top of particle $i$. This
probability can be extracted from the dimensionless pair correlation
function $g(r)$ or its Fourier transform $S(q)$: $P_{\times }=
\int d(q/q_m)[S(q/q_m)-1]$.

For the vortex case the quantity $n\lambda $ of the above
one--dimensional calculation is identified with $n\xi ^2$, $\xi $
being the core radius. Finally, if in our equation for the resistivity
$\rho _{ab}=(\phi _0/c)^2nD/k_BT=(\phi _0/c)^2P_{\times }/(\eta _0\xi
^2)$ we replace the expression for the Bardeen--Stephen viscosity,
$\eta _0=(\hbar /e)^2(\pi /2)/(\rho _{abN}\xi ^2)$ with $\rho _{abN}$
the normal state resistivity , we obtain for the resistivity jump at
$T_m$

\begin{equation}
{\Delta \rho _{ab}(T_m)\over\rho _{abN}} = P_{\times }[S(q/q_m)],
\label{jum} \end{equation}
where $P_{\times }[S(q/q_m)]$ means that $P_{\times }$ is a functional of
$S(q/q_m)$ and is universal at melting.
 This equation is analogous to Eq.(\ref{roc}) and constitutes
the main result of this work. 

The  calculation of the probability $P_{\times }$ of finding vortices in the cutting
configuration requires the knowledge of the of the  exact correlation 
functions $g(r)$. An estimate can be extracted from the 
calculations of the ``cage" model described in Ref.\cite{crabtree}. Within that simple 
an entanglement length $l_z$ is defined as the distance along the field required for
a vortex to diffuse a distance of the order of the vortex--vortex separation $a_0$:
$l_z \sim g\phi_0/(k_BT B)$, with $g$ the tilt modulus, and $a_0^2 \sim \phi_0/B$. 
From the above discussion, we estimate 
\begin{equation}
P_\times \sim a_0/l_z={k_BT_m \over g} \left({B\over \phi_0}\right)^{1/2}.
\label{times2}
\end{equation} 

On the other hand, within the cage model, a melting temperature can be extracted 
by applying Lindemann's criteron on the solid phase\cite{crabtree}:
$k_B T_m =c_L^2 \varepsilon_0 (m_{\bot}/m_z)^{1/2} ( \phi_0/B)^{1/2}$,
with $m_{\bot}$ and $m_z$ being respectively the  in--plane effective mass 
and the out--of--plane effective mass, and $\epsilon_0 = (m_z/m_{\bot}) g$ 
a coupling constant giving the  interaction energy per unit lenth for the vortex lines. 
Also, $c_L$ is the Lindemann constant, which for vortices is of order 
$0.1$--$0.2$\cite{blatter}.
Collecting this with Eq.(\ref{times2}) we obtain 
\begin{equation} 
P_\times \sim c_L^2 (m_{\bot}/m_z)^{1/2}.
\end{equation}

If we assume that the anisotropy on the effective mass is the same as that
of the normal resistivities, we obtain $\Delta \rho (T_m)/\rho _N\sim 10^{-2}$
for YBaCuO and $\Delta \rho (T_m)/\rho _N\sim 10^{-4}$  for 
BISCO, which should be compared with the jumps observed
in experiments of
$0.1$\cite{safar} and $10^{-4}$\cite{fuchs} respectively. 


An alternative estimate of the relative jump at melting
can be can be computed from the expression $P_{\times}=\exp{-U_\times/k_BT_m}$,
and using results from experimental fits\cite{fendrich} and numerical 
simulations\cite{wilkin} that indicate $U_\times \sim k_BT_m$. 
For the observed    $\Delta \rho (T_m)/\rho _N=0.1$ we obtain
 $U_\times = 2.3k_BT_m$, whereas the numerical simulations give 
  $U_\times = 7.5 k_BT_m$. Even though there is a numerical 
  discrepancy, the fact that the crossing energy is independent of
  field at the melting temperature constitutes, in our  view, an 
  additional element supporting the applicability of Verlet's 
  criterion in the vortex state.
  
Our theory asumes a continuum approximation vor the vortex degrees of freedom,
and will be valid as long as $l_z$ is larger than the interplane separation $s$.
When $l_z$ becomes of the order of $s$ one attains the so called decoupled 
or ``superentangled" regime,
in which our approximations are no longer valid.  Therefore, we expect deviations 
from the universal jump in the resistance for magnetic fields larger than
$B_{x2}= g\phi_0/sk_BT$.



An experimental proof of the universalities in $S(q)$ could be obtained
from neutron scattering experiments\cite{cubitt} performed at the melting temperature,
or probably by decorating a sample rapidly quenched from the melting point.

In summary,
 we have presented an explanation for the universal jump in the resistivity
at the melting transition of vortices in high temperature superconductors.
The theory is based on universalities at the melting transition as reflected in 
the structure factor, together with vortex cutting dominating the  viscosity 
of vortices.

\acknowledgments
A. G. R. acknowledges partial support from the National Science
Foundation. We acknowledge conversations with Charlie
Doering, Mark Dykman, Franco Nori, Hugo Safar and
Len Sander. We thank  V. Vinokur for very usefull
remarks.



\begin{references}
\bibitem{crabtree}  G. Crabtree and D. Nelson, Physics Today {\bf 50},
38, April 1997.

\bibitem{blatter} G. Blatter {\it et al.}, Rev. Mod. Phys. {\bf 66} 1125 (1994).
 

\bibitem{safar}  H. Safar {\it et al.} Phys. Rev. Lett. {\bf 69} 824
(1992); W. K. Kwok {\it et al.} Phys. Rev. Lett. {\bf 69} 3370 (1993).

\bibitem{bariloche}  H. Pastoriza, M. F. Goffman, A. Arribere, F. de
la Cruz, Phys. Rev. Lett. {\bf 72} 2951 (1994).

\bibitem{hansen}  J.-P. Hansen and L. Verlet, Phys. Rev. {\bf 184},
159 (1969).

\bibitem{tvr}  T. V. Ramakrishnan and M. Yussouf Phys. Rev. {\bf 18},
2775 (1979).

\bibitem{ziman}  J. M. Ziman, Principles of the Theory of Solids,
Cambridge University Press (1972) Chapter 2.

\bibitem{lowen}  H. L\"{o}wen, T. Palberg and R. Simon, Phys. Rev.
Lett. {\bf 70}, 1557 (1993).

\bibitem{indrani}  A. V. Indrani and S. Ramaswamy, Phys. Rev. Lett.
{\bf 73} 360 (1994).

\bibitem{medina}  M. Medina-Noyola, Phys. Rev. Lett {\bf 60} 2705
(1988).

\bibitem{tinkham}  M. Tinkham, Introduction to Superconductivity, Mc
Graw Hill, New York, (1996) p.163.

\bibitem{bs}  J. Bardeen and M. J. Stephen, Phys. Rev. {\bf 140},
A1197 (1965). 

\bibitem{jagla}  E. Jagla and C. Balseiro, Phys. Rev. Lett. {\bf 77} ,
1588 (1996).

\bibitem{pastoriza}  D. L\'{o}pez {\it et al.}, Phys. Rev. B {\bf 53},
8895 (1996). D. L\'{o}pez, E. F. Righi, G, Nieva, F. de la Cruz, Phys.
Rev. Lett. {\bf 76}, 4034 (1996).

\bibitem{nelson_cut}  M. C. Marchetti and D. Nelson, Physica C {\bf
174}, 40 (1991), S. P. Obukov and M. Rubenstein, Phys. Rev. Lett. {\bf
65}, 1279 (1990).

\bibitem{risken}  ``The Fokker--Planck equation'', by H. Risken,
Ch.11. Sringer Verlag (1984).

\bibitem{sengupta}  S. Sengupta {\it et. al} Phys. Rev. Lett. {\bf
67}, 3444 (1991). 

\bibitem{cubitt} R. Cubitt {\it et al.}, Nature {\bf 365}, 407 (1993).

\bibitem{fuchs} D. T. Fuchs {\it et al.}, Phys. Rev. B {\bf 54} R796 (1996).

\bibitem{fendrich} J. A. Fendirch {\it et al.}, Phys. Rev. Lett. {\bf
74}, 1210 (1995).

\bibitem{wilkin} N. K. Wilkin and M. A. Moore, Physica C {\bf 325-240} 
2641 (1994). 
\end{references}
\end{document}